\documentclass[
 reprint,
 amsmath,amssymb,
 aps,
]{revtex4-2}

\usepackage{graphicx}
\usepackage{dcolumn}
\usepackage{bm}
\usepackage{epstopdf}

\begin{document}

\preprint{APS/123-QED}

\title{Measurement of hyperfine constants and the isotope shift of rubidium 5P$_{1/2}$ excited-state using Saturated Absorption Spectroscopy}
\author{P.M. Rupasinghe, Fiona Wee, Thomas Bullock, Jiaxing Liu}%
\affiliation{Department of Physics, State University of New York at Oswego, Oswego, NY 13126}




\date{\today}

\begin{abstract}
The Saturated Absorption Spectroscopy (SAS) was performed to measure the hyperfine energy splittings of rubidium 5P$_{1/2}$ excited state using a homemade external-cavity diode laser (ECDL) operating at 795 nm. Any nonlinearities associated with ECDL scans were removed by using a low-expansion confocal Fabry-Perot cavity and hence created a linearized frequency axis for the spectra collected in a fully automated fashion. We report our measurements for the magnetic dipole coupling constants 120.79(29) and 407.75(50) for $^{85}Rb$ and $^{87}Rb$ respectively.  
\end{abstract}

\maketitle



\section{Introduction}
Precise measurements of hyperfine structure and isotope shifts of multi-electron atomic systems provide valuable insight into the nuclear properties as well as our understanding of the electronic wavefunctions near the nucleus. Therefore, they play an important role in the development of new atomic theories that may rely on various approximation techniques. For example, two different theoretical methods estimate $^{85}$Rb magnetic-dipole coupling constants for the 5P$_{1/2}$ state as  69.8 MHz and 120.4 MHz while the later being close to the experimental results\cite{Safranova1999}. 

Rubidium can be considered as one of well studied alkali systems with wide range of applications in frequency stabilization techniques, laser cooling and trapping, atomic clocks, development of atomic sensors, and quantum computing. A natural sample of rubidium contains two isotopes $^{85}Rb$ and $^{87}Rb$ with 70$\%$ and 30$\%$ abundances respectively. This system is an attractive candidate for experimental study using laser spectroscopy techniques primarily due to convenience in accessibility of D1 (5S$_{1/2}$$\rightarrow$5P$_{1/2}$) and D2 (5S$_{1/2}$$\rightarrow$5P$_{3/2}$) excitation transitions using commonly available laser diodes operating at 795 nm and 780 nm as well as the availability of sufficient vapor pressure at or near the room temperature.
For such an atomic system the energy of the hyperfine level measured from the center-of-mass energy can be written as,
\begin{equation}
	W(F)=\frac{1}{2}AK + B\frac{(3/2)K(K+1)-2I(I+1)J(J+1)}{2I(2I-1)2J(2J-1)}
	\label{HFenergy}
\end{equation} 
where $K = F(F+1)-I(I+1)-J(J+1)$, $A$ and $B$ are the magnetic-dipole and electric-quadrupole hyperfine coupling constants, $I$ is the nuclear spin, $J$ is the total electronic angular momentum, and $F$ is the total angular momentum of the atom. The second term only contributes when $J > \frac{1}{2}$. 

Hyyperfine constants of rubidium 5$P_{1/2}$ excited state have been measured previously with different techniques\cite{Barnergee2004,Barwood1991,beacham1971,Arimondo1977}. Here, we present another measurement of hyperfine coupling constants and the isotope shift of rubidium 5$P_{1/2}$ state using a distinct approach from the previous techniques.  

\section{Experimental Details}
\subsection{Optical System and Experimental Layout}
\begin{figure}[h!]
	\centering
	\includegraphics[width=0.8\linewidth]{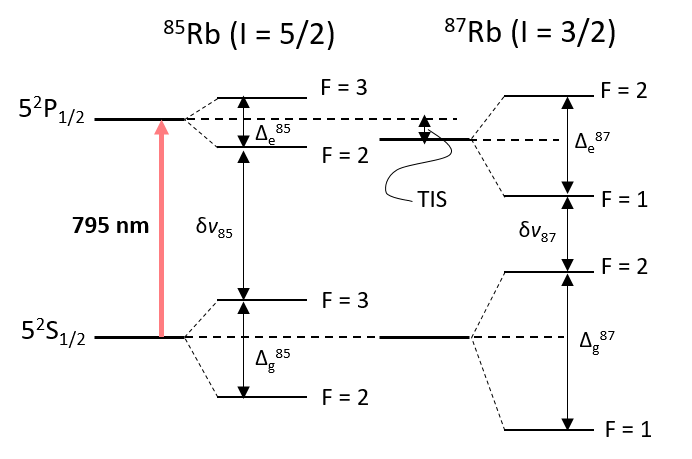}
	\caption{Hyperfine structure of relevant states for two naturally occurring isotopes of rubidium ($^{85}Rb$ ($~70\%$ abundance) $\&$ $^{87}Rb$ (~30$\%$ abundance)). $\Delta^{85}_e$, $\Delta^{87}_e$, and $TIS$ represent hyperfine splitting of $^{85}Rb$, hyperfine splitting of $^{87}Rb$, and the transition isotope shift of the 5P$_{1/2}$ excited state. Similarly, $\Delta^{85}_g$ and $\Delta^{87}_g$ represent hyperfine splittings for the ground state. $\Delta\nu_{85}$ and $\Delta\nu_{87}$ represent the least frequency gaps between excited state and the ground state in $^{85}$Rb and $^{87}$Rb respectively.}
	\label{fig:levelscheme}
\end{figure}

We excited ground state rubidium using 795 nm laser light according to the scheme shown in Figure \ref{fig:levelscheme}.

\begin{figure}[h!]
	\centering
	\includegraphics[width=1.0\linewidth]{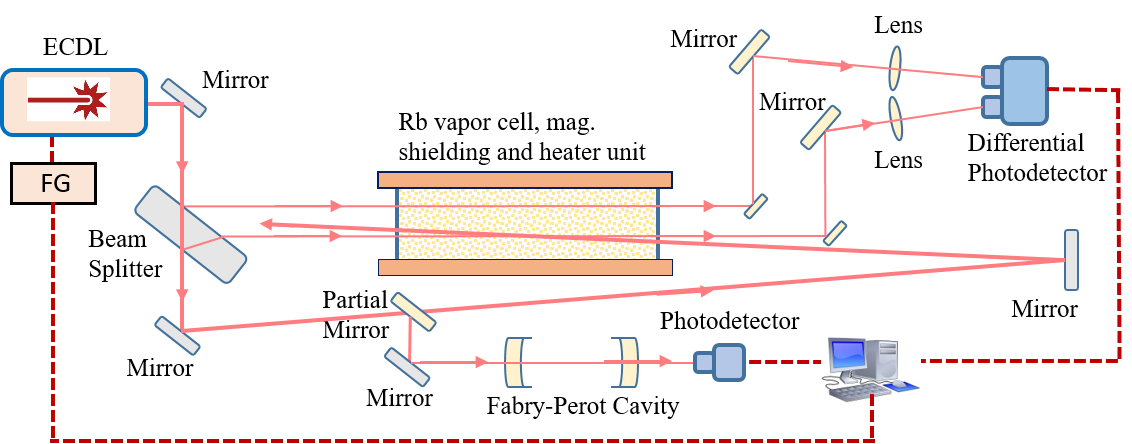}
	\caption{Experimental setup}
	\label{fig:expsetup}
\end{figure}

The external-cavity diode laser (ECDL) used in this experiment is a homemade system based on an existing design\cite{Arnold98, Hawthorn2001} with a 10 mW laser diode. As shown in the Figure \ref{fig:expsetup} two low-power parallel beams with each having power of approximately 25 $\mu$W transmitted through the Rb vapor cell held at 47 $^o$C are directed to a differential photodiode (Thorlabs PDB450A). While the laser is scanning through the relevant transition each beam produces an identical Doppler-broadened direct-absorption signal at individual photodiodes. This corresponds to a nearly zero-signal at the detector output (difference signal). However, when a relatively strong beam (pump-beam) of power approximately 140 $\mu$W is present, a large number of atoms absorb light from the strong beam. As a result, the direct-absorption signal corresponds to the relevant probe-beam shows absorption dips widely known as (hole burning) at certain frequencies.

The hole burning can occur at hyperfine transition frequencies as well as at exactly in the middle of each pair of hyperfine transition frequencies due to the Doppler effect-caused transitions (cross-over resonances). These dip-absorption features show up due to the fact that there are two velocity classes of atoms exist in the cell such that the pump beam is in resonance with one hyperfine transition, and the probe beam is in resonance with the other\cite{Preston1996}. 

When the pump (strong beam) is present the output signal (difference signal) of the differential photodetector shows sharp peaks only correspond to hyperfine as well as cross-over resonances as shown in Figures \ref{fig:Rb85sinletrans} and \ref{fig:Rb87sinletrans}. 

\subsection{Data acquisition system}
After tuning the laser to near transition frequency we scan the laser by applying a triangular voltage signal such that we collect up-scans and down-scans alternatively. During each scan period our labview program samples about 1000 data points per channel in order to create spectra using National Instruments BNC2110 DAQ device. We record hundreds of spectra under various experimental conditions in fully automated fashion with 20-30 scans in each run. Specifically, for each scan we collect the laser scan voltage ramp (up and down), corresponding Fabry-Perot signal, and the vapor cell signal (the saturated absorption spectrum). 

\section{Data analysis and results}
\subsection{Creating a linearized frequency axis}
\begin{figure}[h!]
	\centering
	\includegraphics[width=0.9\linewidth]{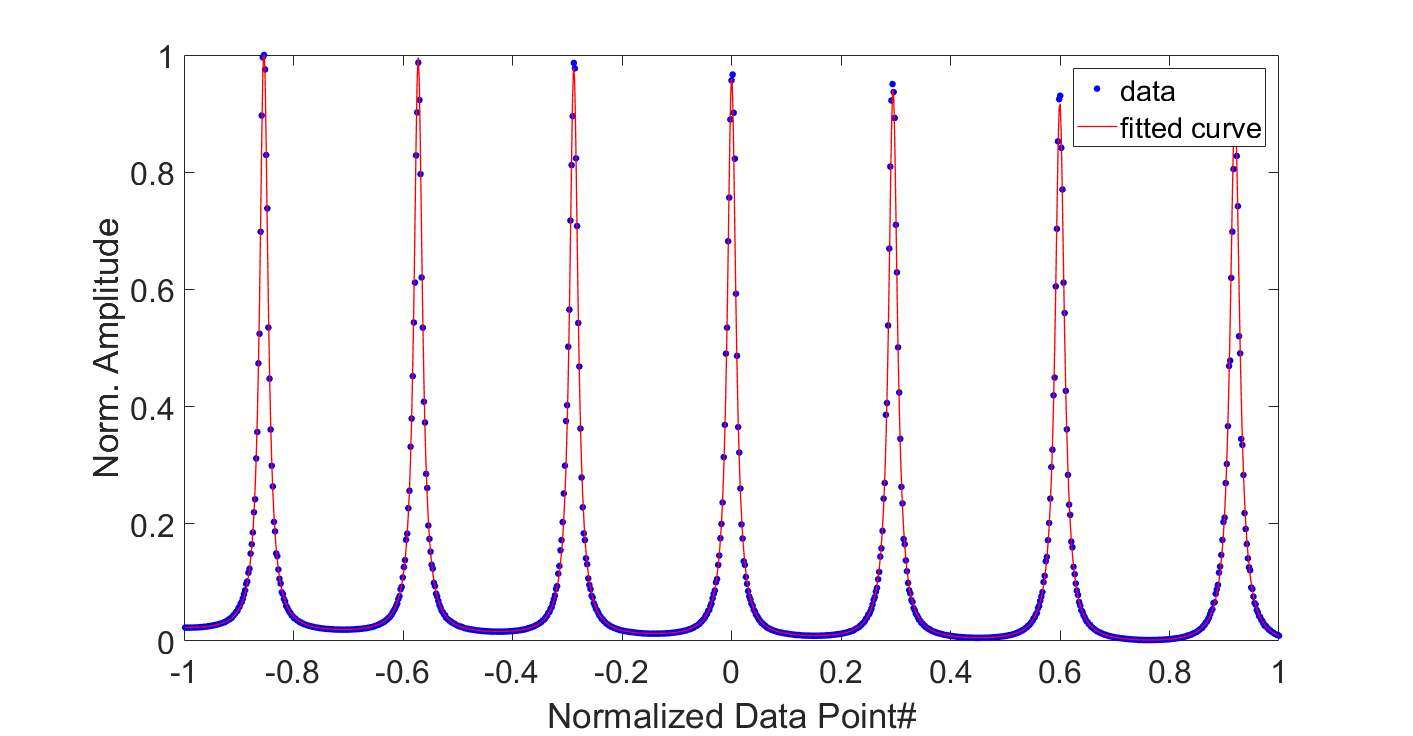} 
	\caption{A sample fit of the Fabry-Perot spectra. Notice the slightly increasing spacing between peaks. In this fit, a 4th order polynomial has been used for f(x).}
	\label{fig:FPfit}
\end{figure}

Data analysis procedure starts with Fabry-Perot spectrum. We fit Fabry-Pérot signal to modified Airy function as a function of normalized data point$\#$ ($x$) as shown in equation \ref{FPmodel}. 
\begin{equation}
	A(x)=\frac{b_0+b_{1} x}{1+F sin^{2}\left((\pi/\Delta)f(x) \right)}
	\label{FPmodel}
\end{equation}
where, $b_0$ and $b_1$ are constants, F is the coefficient of finesse, $\Delta$ is the free-spectral range in MHz, and $f(x)=a_0 + a_{1} x + a_{2} x^2 + ...$ is a polynomial. This fit model allows us not only transform data point numbers to a frequency axis but also remove any nonlinearities associated with the laser scan due to the hysteresis of the PZT device that used to change the angle of the diffraction grating of the ECDL. Such a fit is shown in Figure \ref{fig:FPfit}, notice the slightly increasing separations between peaks in the spectrum which is attributed to a slight nonlinearity of the laser scan.

During the fit all the parameters are floated except the free-spectral range which is kept fixed at its nominal value (500 MHz in our case). Initial conditions for parameters $a_0$ and $a_1$ were set by mapping peak positions $x_i$, where, $i$ is the peak number, to relative frequency values separated by FSR and fitting $f(x_i)$ vs $x_i$ to a linear function $a_0 + a_{1}x$. Higher order coefficients $a_2$, $a_3$,... were set to zero. 
After the first fit all the initial conditions has been updated with new fit parameters and refit for the improved results (Bootstrapping). The final fit parameters $a_0$, $a_1$, ... defines our new frequency axis $f(x)$.

\begin{figure}[h!]
	\centering
	\includegraphics[width=0.9\linewidth]{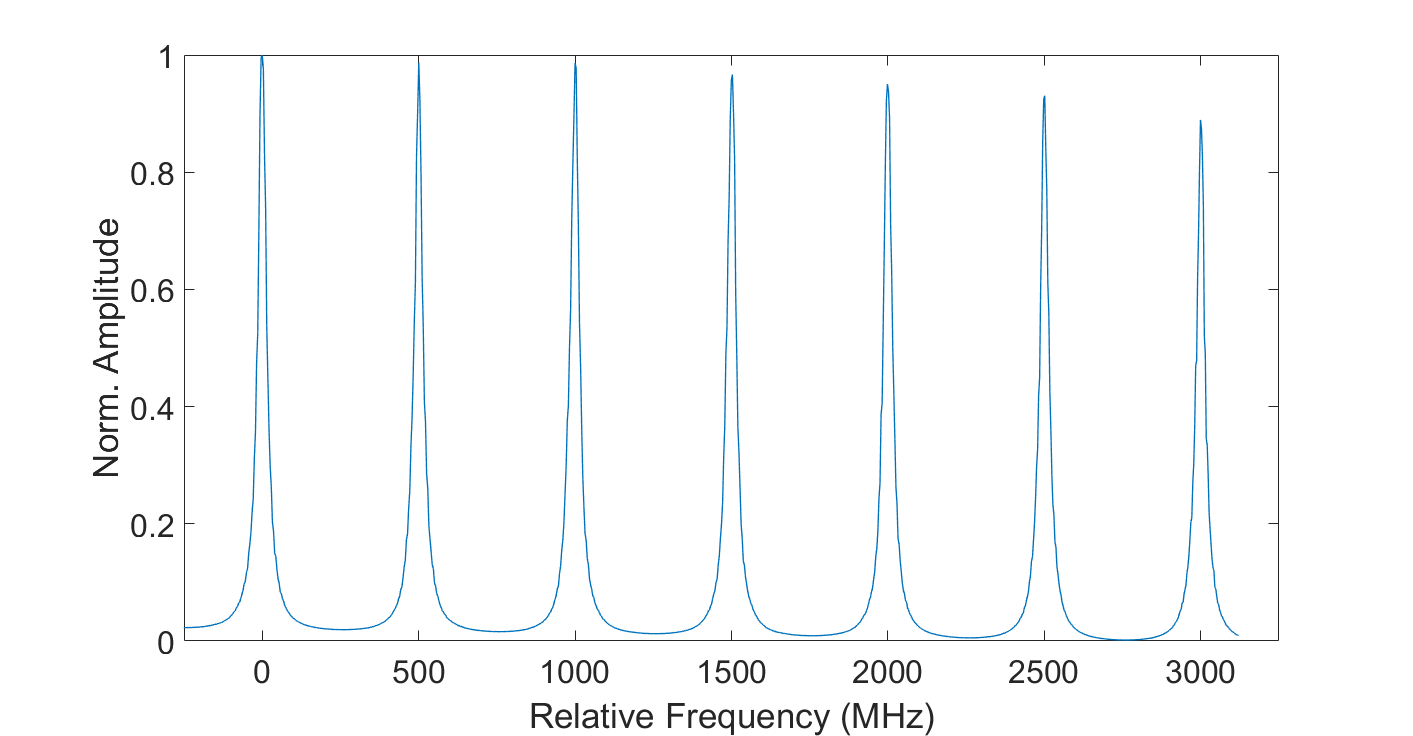}
	\caption{Fabry-Perot spectra is plotted against the new frequency axis $f(x)$. Notice the almost equal spacing between peaks.}
	\label{fig:FPfitFreqDomain}
\end{figure}

Figure \ref{fig:FPfitFreqDomain} shows Fabry-Perot signal plotted against new frequency axis $f(x)$ with a 4$^{th}$ order polynomial. Unlike the Figure \ref{fig:FPfit}, in Figure \ref{fig:FPfitFreqDomain} equal spacing between peaks indicates that the non-linearity associated with laser scans has been removed. There were no statistically significant change of results was observed by using higher order polynomials which further assures that the 4$^{th}$ order polynomial used in this analysis was sufficient to completely remove the nonlinearity. This procedure has been used to create a linearized frequency axes for atomic spectra produced by ECDL scans in many occasions in recent years\cite{Ranjit2013, Ranjit2014, Augenbraun2016}.      

Next we plot the saturated absorption data against the new frequency axis $f(x)$ and fit them to a model function containing sum of Lorentzian line-shapes. Samples of these fits are shown in figures \ref{fig:Rb85sinletrans}, \ref{fig:Rb87sinletrans}, and \ref{fig:twoiso} for $^{85}$Rb, $^{87}$Rb, and both isotope spectra respectively.   

\subsection{Single isotope spectra analysis}
\begin{figure}[h!]
	\centering
	\includegraphics[width=0.9\linewidth]{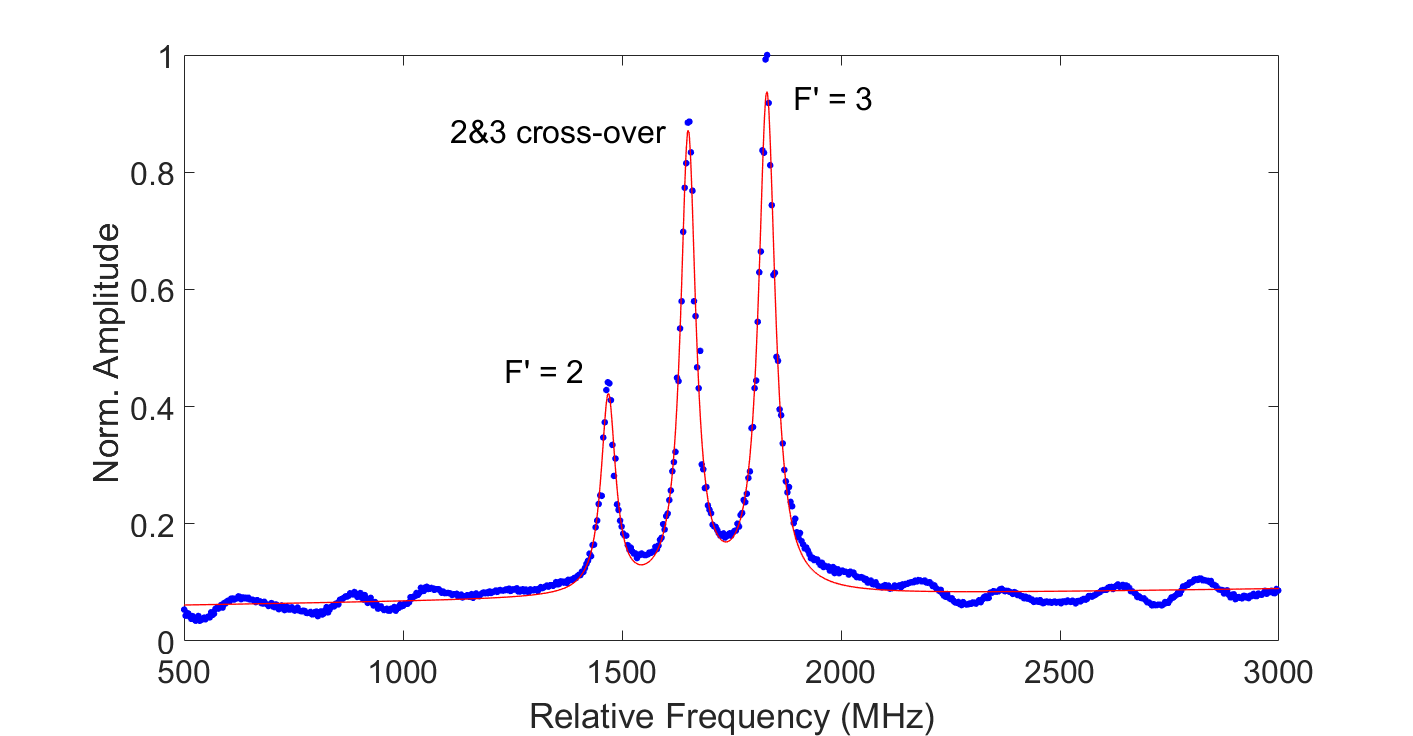}
	\caption{$^{85}$Rb single transition spectrum in frequency domain. An upper transition spectrum corresponds to (F = 3 $\rightarrow$ F' = 2, 3) is shown.}
	\label{fig:Rb85sinletrans}
\end{figure}

\begin{figure}[h!]
	\centering
	\includegraphics[width=0.9\linewidth]{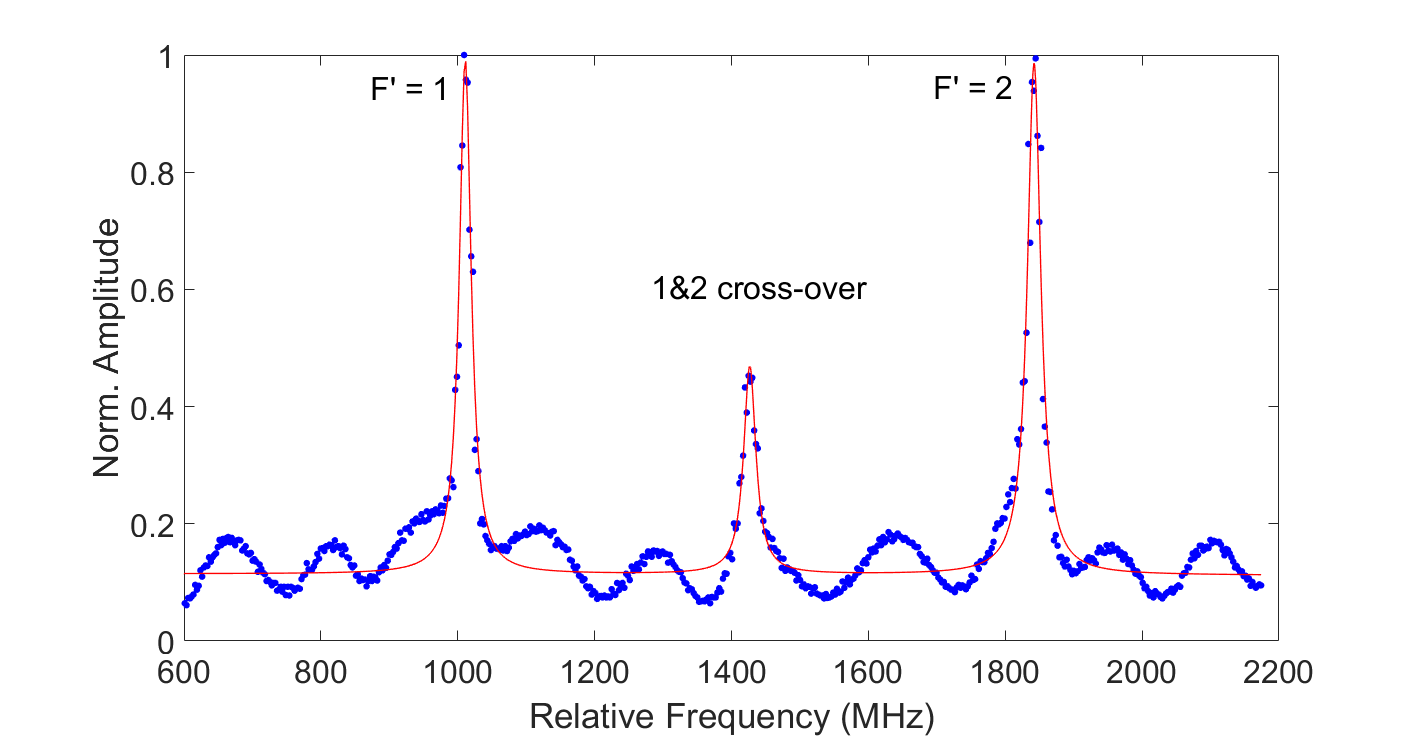}
	\caption{$^{87}$Rb single transition spectrum in frequency domain. An upper transition spectrum corresponds to (F = 2 $\rightarrow$ F' = 1, 2) is shown.}
	\label{fig:Rb87sinletrans}
\end{figure}

Hyperfine splitting values for $^{85}$Rb and $^{87}$Rb spectra shown in figures \ref{fig:Rb85sinletrans} and \ref{fig:Rb87sinletrans} are derived in three independent ways. $\nu_3$-$\nu_1$ (direct), 2($\nu_2$-$\nu_1$)(indirect), and 2($\nu_3$-$\nu_2$)(indirect), where, $\nu_1$, $\nu_2$, and $\nu_2$ are first, second, and third peak positions respectively. These direct and indirect values are a good systematic error check for our frequency axis. For each spectrum, direct and indirect methods yielded similar results reassuring that our frequency axis linearization is done correctly. 

\begin{figure}[h!]
	\centering
	\includegraphics[width=0.9\linewidth]{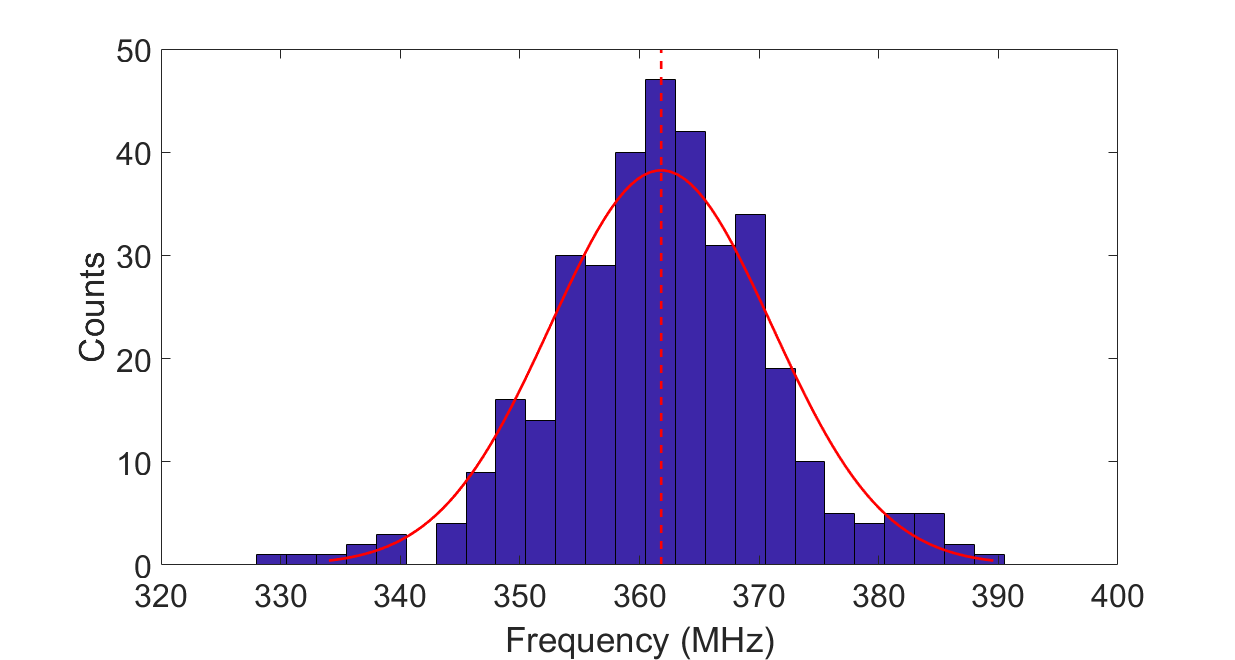}
	\caption{Histogram of the $^{85}$Rb hyperfine splitting measurements extracted from several hundred of scans. The red curve shows the fitted Gaussian function.}
	\label{fig:Rb85_HFS_histo}
\end{figure}

Figure \ref{fig:Rb85_HFS_histo} shows a histogram of $^{85}$Rb hyperfine splitting values extracted in this manner. Similar behavior is observed for $^{87}$Rb splittings as well as isotope shift measurements. 

\subsection{Dual isotope spectra analysis}
\begin{figure}[h!]
	\centering
	\includegraphics[width=0.9\linewidth]{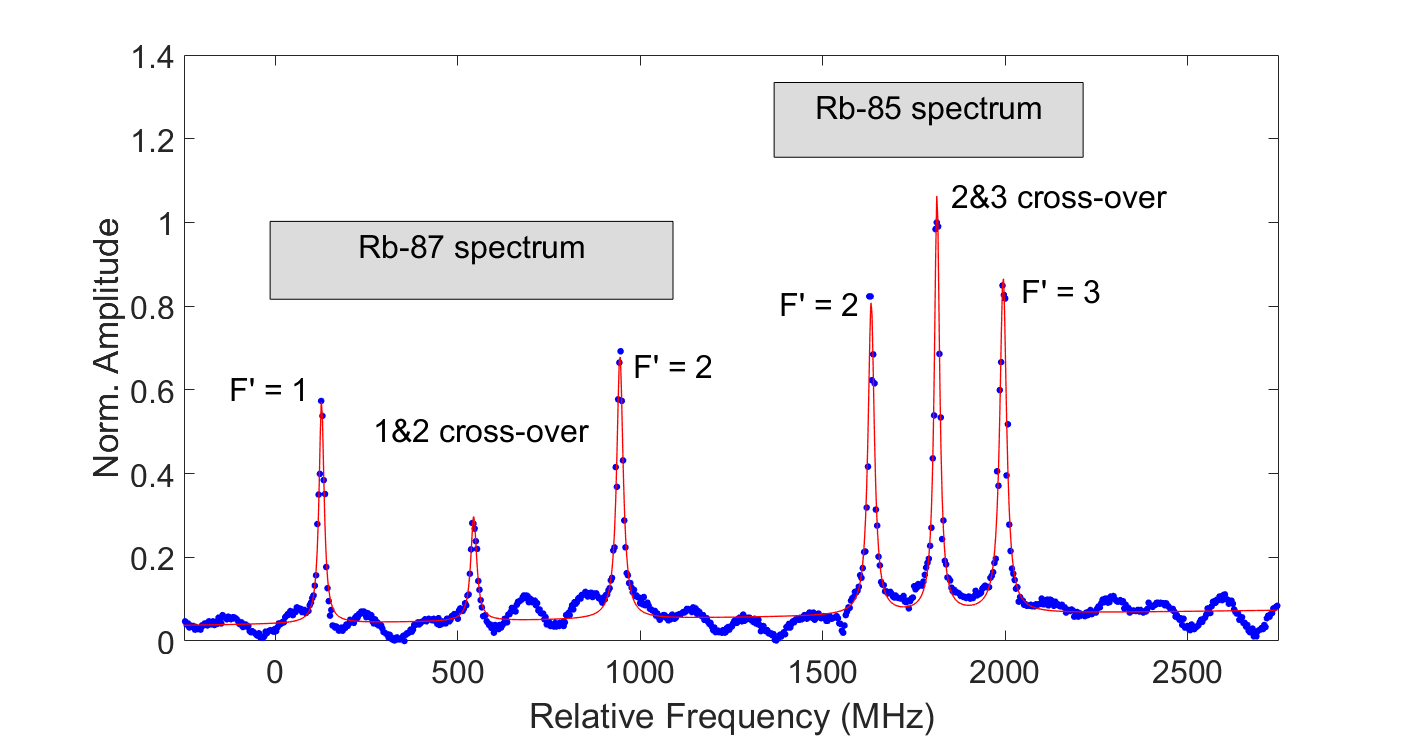}
	\caption{Fitted saturated absorption dual isotope spectrum in frequency domain.}
	\label{fig:twoiso}
\end{figure}

The dual isotope spectrum shown in figure \ref{fig:twoiso} can be used to determine hyperfine splittings for individual isotopes as well as the transition isotope shift between them. Using this spectrum individual hyperfine splittings are extracted using similar analysis done for the single isotope spectra. According the level diagram described in the figure \ref{fig:levelscheme}, transition isotope shift can be determined by the following equation.     
\begin{equation}
	IS = \left(\frac{5}{12}\Delta^{85}_g + \frac{7}{12}\Delta^{85}_e\right) - \left(\frac{3}{8}\Delta^{87}_g  + \frac{5}{8}\Delta^{87}_e\right) + \left(\delta\nu^{85} -\delta\nu^{87}\right)
\end{equation}   
Here, $\left(\delta\nu^{85} -\delta\nu^{87}\right)$ term can be identified as the difference between 1$^{st}$ and 4$^{th}$ peak in the dual isotope spectrum shown in the figure \ref{fig:twoiso}. Similar to hyperfine splitting values, here we can extract the isotope shift values in several independent ways. For example, the term $\left(\delta\nu^{85} -\delta\nu^{87}\right)$ can be determined using various peak positions with the aid of $\Delta^{85}_e$ and $\Delta^{87}_e$. The ground state splittings $\Delta^{85}_g$ and $\Delta^{87}_g$ are taken from \cite{Salomon1999} and \cite{Arimondo1977} respectively. 
 
\subsection{Systematic error search and calibration}
\begin{figure}[h!]
	\centering
	\includegraphics[width=0.9\linewidth]{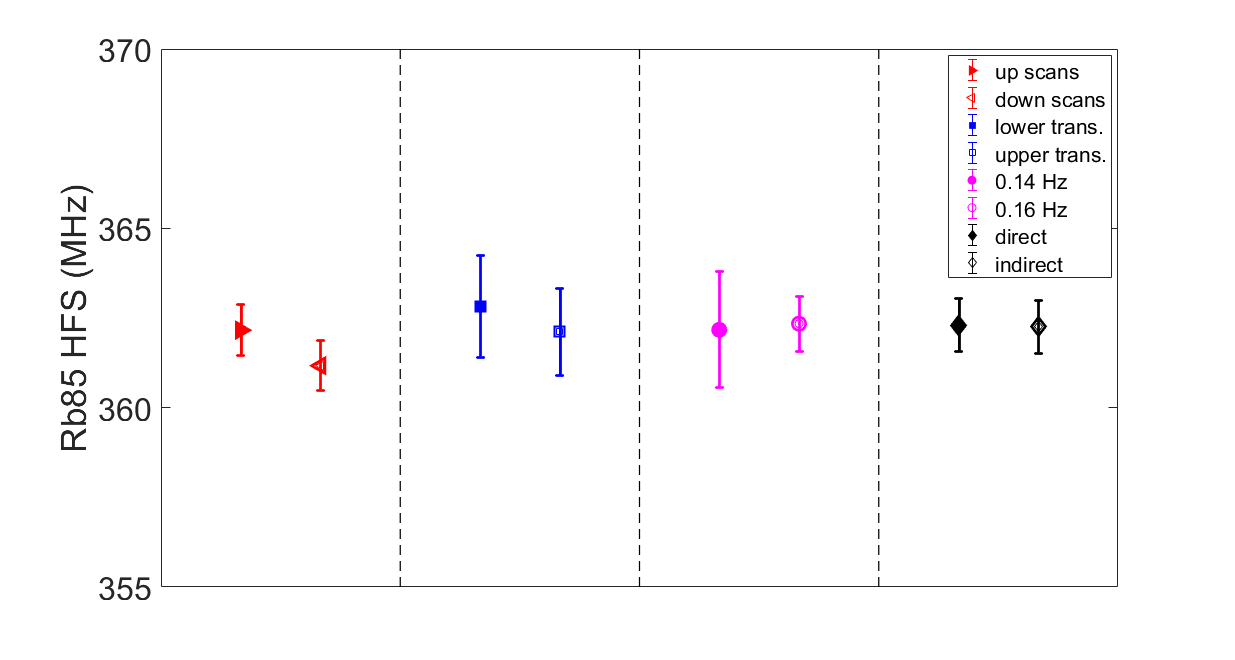}
	\caption{Plot of values obtained for hyperfine splitting of $^{85}Rb$ based on laser scan direction,  scan speed, and excitation scheme (F=2 $\rightarrow$ F'= 2,3) vs (F=3 $\rightarrow$ F'= 2,3).}
	\label{fig:Rb85_systematics}
\end{figure}

In order to assess any potential systematic effects occurred during the course of experiment we divided our data files into different categories and determine the results of our measurements based on laser sweep direction (up scan vs down scan), scan speed, excitation transition, and scan linearization (direct vs indirect measurements). Such a analysis conducted for $^{85}Rb$ hyperfine splitting is shown in the figure \ref{fig:Rb85_systematics}. The plot suggests each subset of data converge to a similar value within their statistical uncertainty. Similarly, we found no evidence of any bias or trends associated with our subset measurements.

During the Fabry-Perot data fitting described in III(a), we used the nominal value of 500 MHz as cavity's FSR. However, this value does not guarantee the actual FSR value present during the experiment. In order to ensure the frequency calibration is done correctly, we measured the FSR few times during the course of the experiment which spanned few weeks to a month. All of our FSR measurements resulted similar results and found the final value of 500.45 MHz. Based on this value we corrected all of our final (uncalibrated) measurements by multiplying a correction factor defines as $c_{cal}$ = 500.45/500.00 = 1.0009. 

\subsection{Results}
Our final (calibrated) hyperfine splittings and isotope shift results are summarized in Table \ref{hfs_table}. 

\begin{table}[h!]
		\caption{Hyperfine splittings and isotope shift measurements}
\begin{tabular}{l l l l}
	\hline \hline \\
Hyperfine Splitting	& $\Delta^{85}_e$ & $\Delta^{87}_e$ & $TIS$\\
	\hline
Final result (MHz)	&362.37  & 815.49 &77.56\\

Stat. error (MHz)	&0.39  & 0.75 & 0.51\\

{\bf Systematic error source:}	&  &  &\\

Scan direction (up vs down)	&0.67  & 0.61 & 0.85\\

Scan speed	&0.10  & 0.16 & 0.24\\

Excitation scheme	& 0.36  & - & -\\

Scan linearization	& 0.02  & 0.23 & 0.16\\

Combined error (MHz)	& 0.86  & 1.00 & 1.03\\
	\hline \hline
\end{tabular}
	\label{hfs_table}
\end{table}

The resulted hyperfine $A$ (magnetic-dipole) coupling constants are listed in Table \ref{A_table} along with previous measurements for comparison.  

\begin{table}[h!]
	\caption{Magnetic-dipole coupling constant measurements}
	\begin{tabular}{l l l}
		\hline \hline \\
		Source	& $^{85}$Rb & $^{87}$Rb\\
		\hline
		This work (MHz)	&120.79(29)  & 407.75(50)\\
		Ref. \cite{Barnergee2004} (MHz)	&120.640(20)  & 406.147(15)\\
		Ref. \cite{Barwood1991} (MHz)	&120.499(10)  & 408.328(15)\\
		Ref. \cite{Arimondo1977} (MHz)	&120.72(25)  & 406.2(8)\\
		\hline \hline
	\end{tabular}
	\label{A_table}
\end{table}

\section{Concluding remarks}
Through saturated absorption spectroscopy (SAS), the hyperfine splittings of Rb 5P$_{1/2}$ excited state have been measured using a homemade ECDL operating at 795 nm. Using these splittings hyperfine A coupling constants for both isotopes have been extracted. Our $^{85}$Rb value shows an agreement with previous measurements. However, our $^{87}$Rb value shows a slight disagreement with previous measurements. It is also important to note that previous measurements described in \cite{Barnergee2004} and \cite{Barwood1991} slightly disagree with the values obtained for both isotopes. We believe our measurements will add a significant contribution to the existing pool of measurements in the literature.   

\begin{acknowledgments} 
The authors would like to thank SUNY-Oswego Office of Research and Sponsored Programs (ORSP) for supporting this work.
\end{acknowledgments}

\bibliography{Rbbiblio}

\providecommand{\noopsort}[1]{}\providecommand{\singleletter}[1]{#1}%
\begin{thebibliography}{12}%
\makeatletter
\providecommand \@ifxundefined [1]{%
 \@ifx{#1\undefined}
}%
\providecommand \@ifnum [1]{%
 \ifnum #1\expandafter \@firstoftwo
 \else \expandafter \@secondoftwo
 \fi
}%
\providecommand \@ifx [1]{%
 \ifx #1\expandafter \@firstoftwo
 \else \expandafter \@secondoftwo
 \fi
}%
\providecommand \natexlab [1]{#1}%
\providecommand \enquote  [1]{``#1''}%
\providecommand \bibnamefont  [1]{#1}%
\providecommand \bibfnamefont [1]{#1}%
\providecommand \citenamefont [1]{#1}%
\providecommand \href@noop [0]{\@secondoftwo}%
\providecommand \href [0]{\begingroup \@sanitize@url \@href}%
\providecommand \@href[1]{\@@startlink{#1}\@@href}%
\providecommand \@@href[1]{\endgroup#1\@@endlink}%
\providecommand \@sanitize@url [0]{\catcode `\\12\catcode `\$12\catcode
  `\&12\catcode `\#12\catcode `\^12\catcode `\_12\catcode `\%12\relax}%
\providecommand \@@startlink[1]{}%
\providecommand \@@endlink[0]{}%
\providecommand \url  [0]{\begingroup\@sanitize@url \@url }%
\providecommand \@url [1]{\endgroup\@href {#1}{\urlprefix }}%
\providecommand \urlprefix  [0]{URL }%
\providecommand \Eprint [0]{\href }%
\providecommand \doibase [0]{https://doi.org/}%
\providecommand \selectlanguage [0]{\@gobble}%
\providecommand \bibinfo  [0]{\@secondoftwo}%
\providecommand \bibfield  [0]{\@secondoftwo}%
\providecommand \translation [1]{[#1]}%
\providecommand \BibitemOpen [0]{}%
\providecommand \bibitemStop [0]{}%
\providecommand \bibitemNoStop [0]{.\EOS\space}%
\providecommand \EOS [0]{\spacefactor3000\relax}%
\providecommand \BibitemShut  [1]{\csname bibitem#1\endcsname}%
\let\auto@bib@innerbib\@empty
\bibitem [{\citenamefont {Safronova}\ \emph {et~al.}(1999)\citenamefont
  {Safronova}, \citenamefont {Johnson},\ and\ \citenamefont
  {Derevianko}}]{Safranova1999}%
  \BibitemOpen
  \bibfield  {author} {\bibinfo {author} {\bibfnamefont {M.~S.}\ \bibnamefont
  {Safronova}}, \bibinfo {author} {\bibfnamefont {W.~R.}\ \bibnamefont
  {Johnson}},\ and\ \bibinfo {author} {\bibfnamefont {A.}~\bibnamefont
  {Derevianko}},\ }\href@noop {} {\bibfield  {journal} {\bibinfo  {journal}
  {Phys.\ Rev. A}\ }\textbf {\bibinfo {volume} {60}},\ \bibinfo {pages} {6}
  (\bibinfo {year} {1999})}\BibitemShut {NoStop}%
\bibitem [{\citenamefont {Barnergee}\ \emph {et~al.}(2004)\citenamefont
  {Barnergee}, \citenamefont {das},\ and\ \citenamefont
  {Natarajan}}]{Barnergee2004}%
  \BibitemOpen
  \bibfield  {author} {\bibinfo {author} {\bibfnamefont {A.}~\bibnamefont
  {Barnergee}}, \bibinfo {author} {\bibfnamefont {D.}~\bibnamefont {das}},\
  and\ \bibinfo {author} {\bibfnamefont {V.}~\bibnamefont {Natarajan}},\
  }\href@noop {} {\bibfield  {journal} {\bibinfo  {journal} {Europhys. Lett.,}\
  }\textbf {\bibinfo {volume} {65}},\ \bibinfo {pages} {172} (\bibinfo {year}
  {2004})}\BibitemShut {NoStop}%
\bibitem [{\citenamefont {Barwood}\ \emph {et~al.}(1991)\citenamefont
  {Barwood}, \citenamefont {Gill},\ and\ \citenamefont {Rowley}}]{Barwood1991}%
  \BibitemOpen
  \bibfield  {author} {\bibinfo {author} {\bibfnamefont {G.~P.}\ \bibnamefont
  {Barwood}}, \bibinfo {author} {\bibfnamefont {P.}~\bibnamefont {Gill}},\ and\
  \bibinfo {author} {\bibfnamefont {W.~R.~C.}\ \bibnamefont {Rowley}},\
  }\href@noop {} {\bibfield  {journal} {\bibinfo  {journal} {Appl. Phys. B,}\
  }\textbf {\bibinfo {volume} {53}},\ \bibinfo {pages} {142} (\bibinfo {year}
  {1991})}\BibitemShut {NoStop}%
\bibitem [{\citenamefont {Beacham}\ and\ \citenamefont
  {Andrew}(1971)}]{beacham1971}%
  \BibitemOpen
  \bibfield  {author} {\bibinfo {author} {\bibfnamefont {J.~R.}\ \bibnamefont
  {Beacham}}\ and\ \bibinfo {author} {\bibfnamefont {K.~L.}\ \bibnamefont
  {Andrew}},\ }\href@noop {} {\bibfield  {journal} {\bibinfo  {journal}
  {Journal of Opt. Soc. of America}\ }\textbf {\bibinfo {volume} {61}},\
  \bibinfo {pages} {2} (\bibinfo {year} {1971})}\BibitemShut {NoStop}%
\bibitem [{\citenamefont {Arimondo}\ \emph {et~al.}(1977)\citenamefont
  {Arimondo}, \citenamefont {Inguscio},\ and\ \citenamefont
  {Violino}}]{Arimondo1977}%
  \BibitemOpen
  \bibfield  {author} {\bibinfo {author} {\bibfnamefont {E.}~\bibnamefont
  {Arimondo}}, \bibinfo {author} {\bibfnamefont {M.}~\bibnamefont {Inguscio}},\
  and\ \bibinfo {author} {\bibfnamefont {P.}~\bibnamefont {Violino}},\
  }\href@noop {} {\bibfield  {journal} {\bibinfo  {journal} {Reviews of Modern
  Physics,}\ }\textbf {\bibinfo {volume} {49}},\ \bibinfo {pages} {1} (\bibinfo
  {year} {1977})}\BibitemShut {NoStop}%
\bibitem [{\citenamefont {Arnold}\ \emph {et~al.}(1998)\citenamefont {Arnold},
  \citenamefont {Wilson},\ and\ \citenamefont {Boshier}}]{Arnold98}%
  \BibitemOpen
  \bibfield  {author} {\bibinfo {author} {\bibfnamefont {A.~S.}\ \bibnamefont
  {Arnold}}, \bibinfo {author} {\bibfnamefont {J.~S.}\ \bibnamefont {Wilson}},\
  and\ \bibinfo {author} {\bibfnamefont {M.~G.}\ \bibnamefont {Boshier}},\
  }\href@noop {} {\bibfield  {journal} {\bibinfo  {journal} {Rev. Sci.
  Instrum.}\ }\textbf {\bibinfo {volume} {69}},\ \bibinfo {pages} {1236}
  (\bibinfo {year} {1998})}\BibitemShut {NoStop}%
\bibitem [{\citenamefont {Hawthorn}\ \emph {et~al.}(2001)\citenamefont
  {Hawthorn}, \citenamefont {Weber},\ and\ \citenamefont
  {Scholtena}}]{Hawthorn2001}%
  \BibitemOpen
  \bibfield  {author} {\bibinfo {author} {\bibfnamefont {C.~J.}\ \bibnamefont
  {Hawthorn}}, \bibinfo {author} {\bibfnamefont {K.~P.}\ \bibnamefont
  {Weber}},\ and\ \bibinfo {author} {\bibfnamefont {R.~E.}\ \bibnamefont
  {Scholtena}},\ }\href@noop {} {\bibfield  {journal} {\bibinfo  {journal}
  {Rev. Sci. Instrum.}\ }\textbf {\bibinfo {volume} {72}},\ \bibinfo {pages}
  {12} (\bibinfo {year} {2001})}\BibitemShut {NoStop}%
\bibitem [{\citenamefont {Preston}(1996)}]{Preston1996}%
  \BibitemOpen
  \bibfield  {author} {\bibinfo {author} {\bibfnamefont {D.~W.}\ \bibnamefont
  {Preston}},\ }\href@noop {} {\bibfield  {journal} {\bibinfo  {journal} {Am.
  J. Physics,}\ }\textbf {\bibinfo {volume} {64}},\ \bibinfo {pages} {11}
  (\bibinfo {year} {1996})}\BibitemShut {NoStop}%
\bibitem [{\citenamefont {Ranjit}\ \emph {et~al.}(2013)\citenamefont {Ranjit},
  \citenamefont {Schine}, \citenamefont {Lorenzo}, \citenamefont {Schneider},\
  and\ \citenamefont {Majumder}}]{Ranjit2013}%
  \BibitemOpen
  \bibfield  {author} {\bibinfo {author} {\bibfnamefont {G.}~\bibnamefont
  {Ranjit}}, \bibinfo {author} {\bibfnamefont {N.~A.}\ \bibnamefont {Schine}},
  \bibinfo {author} {\bibfnamefont {A.~T.}\ \bibnamefont {Lorenzo}}, \bibinfo
  {author} {\bibfnamefont {A.~E.}\ \bibnamefont {Schneider}},\ and\ \bibinfo
  {author} {\bibfnamefont {P.~K.}\ \bibnamefont {Majumder}},\ }\href@noop {}
  {\bibfield  {journal} {\bibinfo  {journal} {Phys.\ Rev. A}\ }\textbf
  {\bibinfo {volume} {87}},\ \bibinfo {pages} {032506} (\bibinfo {year}
  {2013})}\BibitemShut {NoStop}%
\bibitem [{\citenamefont {Ranjit}\ \emph {et~al.}(2014)\citenamefont {Ranjit},
  \citenamefont {Kealhofer}, \citenamefont {Vukasin},\ and\ \citenamefont
  {Majumder}}]{Ranjit2014}%
  \BibitemOpen
  \bibfield  {author} {\bibinfo {author} {\bibfnamefont {G.}~\bibnamefont
  {Ranjit}}, \bibinfo {author} {\bibfnamefont {D.}~\bibnamefont {Kealhofer}},
  \bibinfo {author} {\bibfnamefont {G.}~\bibnamefont {Vukasin}},\ and\ \bibinfo
  {author} {\bibfnamefont {P.~K.}\ \bibnamefont {Majumder}},\ }\href@noop {}
  {\bibfield  {journal} {\bibinfo  {journal} {Phys.\ Rev. A}\ }\textbf
  {\bibinfo {volume} {89}},\ \bibinfo {pages} {012511} (\bibinfo {year}
  {2014})}\BibitemShut {NoStop}%
\bibitem [{\citenamefont {Augenbraun}\ \emph {et~al.}(2016)\citenamefont
  {Augenbraun}, \citenamefont {Carter}, \citenamefont {Rupasinghe},\ and\
  \citenamefont {Majumder}}]{Augenbraun2016}%
  \BibitemOpen
  \bibfield  {author} {\bibinfo {author} {\bibfnamefont {B.~L.}\ \bibnamefont
  {Augenbraun}}, \bibinfo {author} {\bibfnamefont {A.}~\bibnamefont {Carter}},
  \bibinfo {author} {\bibfnamefont {P.~M.}\ \bibnamefont {Rupasinghe}},\ and\
  \bibinfo {author} {\bibfnamefont {P.~K.}\ \bibnamefont {Majumder}},\
  }\href@noop {} {\bibfield  {journal} {\bibinfo  {journal} {Phys.\ Rev. A}\
  }\textbf {\bibinfo {volume} {94}},\ \bibinfo {pages} {022515} (\bibinfo
  {year} {2016})}\BibitemShut {NoStop}%
\bibitem [{\citenamefont {Bize}\ \emph {et~al.}(1999)\citenamefont {Bize},
  \citenamefont {Sortais}, \citenamefont {Santos}, \citenamefont {Mandache},
  \citenamefont {Clairon},\ and\ \citenamefont {Salomon}}]{Salomon1999}%
  \BibitemOpen
  \bibfield  {author} {\bibinfo {author} {\bibfnamefont {S.}~\bibnamefont
  {Bize}}, \bibinfo {author} {\bibfnamefont {Y.}~\bibnamefont {Sortais}},
  \bibinfo {author} {\bibfnamefont {M.~S.}\ \bibnamefont {Santos}}, \bibinfo
  {author} {\bibfnamefont {C.}~\bibnamefont {Mandache}}, \bibinfo {author}
  {\bibfnamefont {A.}~\bibnamefont {Clairon}},\ and\ \bibinfo {author}
  {\bibfnamefont {C.}~\bibnamefont {Salomon}},\ }\href@noop {} {\bibfield
  {journal} {\bibinfo  {journal} {Europhysics Letters}\ }\textbf {\bibinfo
  {volume} {45}},\ \bibinfo {pages} {558} (\bibinfo {year} {1999})}\BibitemShut
  {NoStop}%
\end{thebibliography}%

\end{document}